\documentclass[12pt]{article}
\usepackage{amsmath}
\usepackage{amssymb}
\usepackage{epsfig}

\begin{document}

\title{Properties of atypical graphs from negative complexities}
\author{Olivier Rivoire \\
\\
{\small  Laboratoire de Physique Th\'eorique et Mod\`eles Statistiques}\\
{\small b\^at. 100, Universit\'e Paris-Sud, 91405 Orsay, France}}
\date{\small April 27, 2004}

\maketitle

\abstract{The one-step replica symmetry breaking cavity method is proposed as a new tool to investigate large deviations in random graph ensembles. The procedure hinges on a general connection between negative complexities and probabilities of rare samples in spin glass like models. This relation between large deviations and replica theory is explicited on different models where it is confronted to direct combinatorial calculations.}\\

{\small {\bf PACS:} 02.10.Ox - 05.50.+q - 75.10.Nr}% Combinatorics; graph theory - Lattice theory and statistics - Spin-glass and other random models

\section{Introduction}

In systems with quenched disorder and not too long range interactions, the free energy density is a self-averaging quantity that converges, with probability one, to a {\it typical} value when the size of the system tends to infinity. The difficulty to compute such typical quantities is responsible for the tremendous activity observed during the last 25 years, and has motivated the introduction of two powerful tools, the {\it replica method} and, its equivalent reformulation, the {\it cavity method} \cite{MezardParisi87b}. Significant advances in the quest for setting these methods on firm mathematical grounds have been made recently, leading to the demonstration of a number of predictions of replica theory, but many challenging related problems are still open \cite{Talagrand03}.

Initially restricted to models with quenched disorder defined on complete graphs, the replica and cavity methods have been extended in many different directions, including the study of metastable and non-equilibrium states, frustrated but non disordered systems, and models defined on dilute, finite connectivity, graphs. The progress has recently led, in the context of the cavity method \cite{MezardParisi01}, to predictions for different important and famous optimization problems, like {\sc k-sat} or coloring. These results are conjectured to be exact \cite{MezardParisi02}, and in some cases, including most notably the random assignment problem, they have been confirmed by a rigorous proof \cite{Aldous00}.

In many optimization problems, part of the quenched disorder is encoded in an ensemble of {\it random graphs} \cite{Bollobas01}. A characteristic of random graphs is that, in the limit of infinite sizes, many of their properties, like for instance the existence of a percolating structure, have a threshold behavior \cite{FriedgutKalai96}: if a graph is randomly chosen from some predefined ensemble, it satisfies a given property, like percolation, with probability zero or with probability one, depending on the ensemble. When a property is satisfied {\it asymptotically almost surely} i.e., with probability one in the infinite size limit, we refer to the graphs having this property as the {\it typical} graphs for this property; the rest, having measure zero, are rare, {\it atypical}, graphs. A change from zero to one for the probability of occurrence of a given property, when some parameter characterizing the ensemble is varied, is a geometrical realization of a phase transition. This connection between statistical physics and graph theory has been recognized for along time; however, up to now it has been restricted to typical graphs.

The object of this paper is to point out that the cavity method, initially developed to treat typical samples, can also provide valuable information on atypical samples and, in particular, on atypical graphs. Given a constrained satisfaction problem, a sample is said {\it frustrated} if it is impossible to satisfy all constraints, and otherwise {\it unfrustrated}. The quantity we will evaluate is the probability of typical, unfrustrated, samples in a regime where the problem is typically satisfiable. 

More precisely, we relate, for a given model, the central object of the cavity method, the {\it complexity}, to the rate function describing the large deviation probabilities of the ground state energy. The conjectured relation, stated in Eq.~(\ref{eq:pN}), is motivated by a general, yet heuristic, argument and is explicitly verified in a number of particular cases for which the complexity and large deviations properties can be computed independently. The examples treated include the random energy model (REM) \cite{Derrida81}, the $\pm J$ spin-glass on a Bethe lattice \cite{MezardParisi02} and a class of lattice glass models \cite{BiroliMezard01}, the most famous of which being the hard-core model.

The connection between complexity and large deviations allows us, when the quenched disorder lies in the random graph structure, to extract some probabilities of atypical graphs from the cavity method. We show for instance how the study of the hard-core model provides us with the asymptotic probability for a random graph to be bipartite, that is, bicolorable. Such a result is particularly interesting considering the very small number of techniques available for studying atypical graphs. Remarkably, other though related statistical physics tools have also been shown, very recently, to give access to different atypical graph properties \cite{EngelMonasson03}.

The paper is organized as follows. Random graphs, which will constitute the most interesting instance of quenched disorder, are introduced in Sec.~\ref{sec:randomgraphs}, together with a discussion of their relation to mean-field approximations. Sec.~\ref{sec:complexity} presents, on the example of lattice glass models, the concept of complexity (also called configurational entropy) in the framework of the cavity method. Sec.~\ref{sec:interpretation} introduces the conjecture relating negative complexities to atypical samples. The proposed relation with large deviations is then exactly verified on the REM in Sec.~\ref{sec:rem}, on the spin glass model in Sec.~\ref{sec:spin-glass} and on the hard-core model in Sec.~\ref{sec:hardcore}. For this later model, applications to graph theory are emphasized, and a generalization to more general lattice glass models is given in the appendix. The issue of the validity of the 1-{\sc rsb} approach, which appears questionable for the spin glass and hard-core models, is addressed in Sec.~\ref{sec:stability}. Finally, the conclusion suggests some potential implications.

\section{Random graphs and Bethe lattices}\label{sec:randomgraphs}

A {\it random graph} is a set of {\it vertices} with {\it edges} randomly connecting pairs of them. At fixed number $N$ of vertices, different classes of random graphs can be defined, depending on the rule used to randomly construct the edges. For instance, having the $N(N-1)/2$ edges present with independent probability $\gamma/N$ defines {\it Erd\H{o}s-R\'enyi random graphs}, named in honor of the mathematicians who first introduced them in the 1960's \cite{ErdosRenyi60}. These graphs have the property that, in the limit $N\to\infty$, the number of edges associated to a vertex, called the {\it degree} $d$ of that vertex, obeys a Poissonian distribution, $p_d=\gamma^de^{-\gamma}/d!$. In the last forty years, graph theory has been the object of much activity in probability theory (see e.g. Ref.~\cite{Bollobas01} for an account), and it has found many applications in the study of real complex networks \cite{RekaBarabasi02}. Recent investigations on the structure of social networks or the Internet have focused on classes of random graphs with different degree distributions, like for instance power-laws, $p_d\propto d^{-\tau}$ \cite{NewmanStrogatz01}. Another class consists in fixing the degree $d$ to be the same for all vertices i.e., $p_d=\delta_{d,r}$; this defines the class, hereafter noted $\mathcal{G}_N^{(r)}$, of {\it random $r$-regular graphs} \cite{Wormald99}.

Due to their local homogeneity, random regular graphs are useful as approximations of (non random) regular lattices. The intrinsic difficulties of three-dimensional Euclidean lattices has indeed oriented attention on mean-field models where only local correlations are treated exactly. In turn, such mean-field models are often related to some particular graph (called {\it lattice} when non random) such that the approximation is exact for models defined on it; for instance the Curie-Weiss mean-field approximation, where all correlations are neglected, is associated with completely connected graphs i.e., lattices where every vertex is connected to all other vertices.

A more refined mean-field approximation is the Bethe-Peierls approximation where nearest neighbor correlations are treated exactly while correlations at higher distances are self-consistently taken into account. In this case, people speak of the {\it Bethe lattice} as the structure on which the approximation is exact. Unfortunately, this structure is {\it not} independent of the particular model under consideration. While the interior of a Cayley tree is appropriate when dealing with unfrustated models such as the Ising model (see e.g. Ref.~\cite{Baxter82}), it is no longer adequate for frustrated models such as glass models, due to an oddly defined thermodynamical limit (see e.g. Ref.~\cite{RivoireBiroli04}). It has been proposed \cite{MezardParisi87c} that random regular graphs provides the suitable structure, valid both for unfrustated and frustrated system. Indeed, in the large $N$ limit ($N$ being the number of vertices) a random regular graph is locally tree-like, but statistical physics models defined on it do not suffer from the strong boundary dependence that affects Cayley trees.

One way to implement this idea is to study a problem of statistical physics on a graph generated randomly with uniform distribution from the class $\mathcal{G}_N^{(r)}$ of random regular graphs. This procedure actually amounts to introducing a quenched disorder whose distribution is defined by the ensemble of random graph; this quenched disorder can be handled by means of the {\it cavity method} \cite{MezardParisi01}.

\section{The cavity method}\label{sec:complexity}

Atypical samples show up in the cavity method when computing the {\it complexity}, also called the {\it configurational entropy}. For definiteness, we present this concept and the cavity method on the example of lattice glass models where the quenched disorder is only coming from the underlying spatial structure, which is taken to be a random regular graph (see Ref.~\cite{MezardParisi03} for more details on the cavity method). Lattice glass models aim at modeling the geometrical {\it frustration} that arises in three dimensional vitreous systems where the locally preferred structure, an icosahedron formed by one atom surrounded by 12 other atoms, is incompatible with the global constraint of tiling the entire space (this is due to the 5-fold rotational symmetry of icosahedrons).

Frustration means that some local constraints forbid a global optimum to be obtained as a sum of local optimizations. In lattice glass models, the local geometrical constraints are expressed as the impossibility, for a given particle, to have more than a prescribed number $\ell\geq 0$ of neighboring particles \cite{BiroliMezard01}. For instance, taking $\ell=0$ defines the {\it hard-core model} where a particle on a vertex forbids the presence of another particle on all the neighboring vertices.

In the absence of frustration, the greatest achievable density $\rho_{\rm cp}$, the {\it close-packing} density, is associated with a crystalline order. In the $\ell=0$ case, it is characterized by an alternation of occupied and empty vertices, empty sites never being adjacent, so that $\rho_{\rm cp}=\rho_{\rm cryst}=1/2$; on the contrary, frustration is present as soon as the underlying lattice has an {\it odd cycle} i.e., a loop with an odd number of vertices, in which case there must exist two neighboring empty vertices. Since odd cycles proliferate in typical random regular graphs, models with a $\ell=0$ constraint defined on random regular graphs are expected to have a close-packing density $\rho_{\rm rcp}$ lower than the crystalline density, $\rho_{\rm rcp}<\rho_{\rm cryst}$ (``rcp'' stands for {\it random close-packing}).

The cavity method allows one to evaluate $\rho_{\rm rcp}$ for lattice glass models on random $r$-regular graphs with different constraints $\ell\leq r$; these models are noted LG($k,\ell$) with $r\equiv k+1$ in the following. For some of the parameters $k$, $\ell$, the cavity method indicates that the phase space has a {\it one-step replica symmetry breaking} (1-{\sc rsb}) structure \cite{BiroliMezard01, RivoireBiroli04}. It means that the high density configurations are organized into distinct components, called {\it states} \cite{MezardParisi87b}. This glassy feature is to be contrasted with the {\it replica symmetric} ({\sc rs}) structure of a liquid state, where the dominant configurations belong to an unique cluster of configurations ({\it one} state only). If there are many states which are themselves organized into greater clusters, it is referred as 2-{\sc rsb}; more generally, a $m$-{\sc rsb} structure corresponds to a hierarchy of $m$ families of clusters contained in one another. The $m\to\infty$ limiting structure, as arises for instance in the SK model \cite{MezardParisi87b}, is said to have a {\it full replica symmetry breaking} (full-{\sc rsb}) pattern.

In the 1-{\sc rsb} framework, a state $\alpha$, which represents a cluster of configurations, is associated with a certain density $\rho_\alpha$; different states can have the same or different densities. The {\it complexity} $\Sigma(\rho)$, also called {\it configurational entropy} in the context of glass theories, depends on this phase space organization and gives the number $\mathcal{N}(\rho)$ of states with a given density $\rho$, through the defining relation,
\begin{equation}\label{eq:complexity}
\mathcal{N}(\rho)\equiv\exp[N\Sigma(\rho)].
\end{equation}

An example of complexity curve as obtained from the cavity method is shown in Fig.~\ref{fig:complexity}. The complexity counts states in the same way that the entropy counts configurations; it is therefore natural to introduce a quantity corresponding to a free-energy, noted $\phi(y)$, defined by
\begin{equation}
e^{-Ny\phi(y)}=\int d\rho\mathcal{N}(\rho)e^{y\rho}=\int  d\rho e^{N[\Sigma(\rho)+y\rho]},
\end{equation}
from which the density $\rho$ and the complexity $\Sigma$ can be deduced through the relations \cite{Monasson95}
\begin{equation}
\begin{split}
\Sigma (y)&=y^2\partial_y \phi (y),\\
\rho (y)&=-\partial_y [y\phi (y)].
\end{split}
\end{equation}
The parameter $y$, the counterpart of the (inverse) temperature, is also the slope of the complexity curve, $y=-\partial\Sigma/\partial\rho$. In contrast to the usual temperature, $y$ is not an external parameter and must be chosen self-consistently. Thus, one can argue that the random close-packing density $\rho_{\rm rcp}$ is given at the $y=y^*$ for which $\phi(y)$ is maximal \cite{MezardParisi03}; it corresponds to $\rho(y^*)=\rho_{\rm rcp}$ and $\Sigma(y^*)=\Sigma(\rho_{\rm rcp})=0$. The values of the complexity $\Sigma(\rho)$ for $\rho<\rho_{\rm rcp}$ (see Fig.~\ref{fig:complexity}), are attributed to the presence of many {\it metastable} states, i.e. states $\alpha$ with $\rho_\alpha<\rho_{\rm rcp}$.

\section{Complexity and large deviations}\label{sec:interpretation}

\begin{figure}
%\centering\resizebox{0.4\textwidth}{!}{\epsfig{file=complexity.ps}}
\centering\epsfxsize=8cm \epsfbox{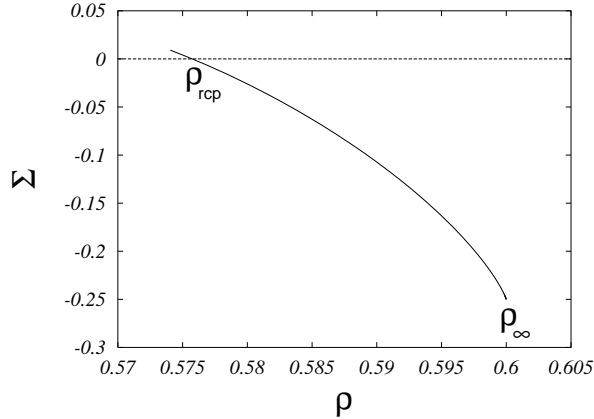}
\caption{\small Complexity curve, as obtained from the 1-{\sc rsb} cavity method, for the LG($k=2,\ell=1$) model. It corresponds to Fig.~6 of Ref.~\cite{RivoireBiroli04} with only the stable part kept and all the negative part for $\rho>\rho_{\rm rcp}\simeq 0.5757$ represented. Note the terminal point of coordinates $\rho_\infty=3/5=0.6$ and $\Sigma_\infty=\frac{4}{5}\ln2-\frac{1}{2}\ln 5\simeq -0.250$, where the slope $-y$ of the curve becomes infinite (see the difference of scale between the two axes).} \label{fig:complexity}
\end{figure}

As shown in Fig.~\ref{fig:complexity}, the {\it negative} part of the complexity curve, on which we will focus here, is instead associated with densities {\it larger} than $\rho_{\rm rcp}$.
This feature is at first sight in contradiction both with the definition of the complexity, Eq.~(\ref{eq:complexity}), which insures that $\Sigma(\rho)\geq 0$, and with the claim that $\rho_{\rm rcp}$ indeed corresponds to a maximum achievable density. 
This paradox is removed if $\mathcal{N}(\rho)\equiv\exp[N\Sigma(\rho)]$ is considered as an average, over the quenched disorder generated by the class of random graphs $G\in\mathcal{G}_N^{(r)}$, of the number of states $\mathcal{N}_G(\rho)\equiv\exp[N\Sigma_G(\rho)]$ for the model on one such graph $G$. By definition $\Sigma_G(\rho)\geq 0$, but, if
\begin{equation}\label{eq:average}
e^{N\Sigma(\rho)}=\frac{1}{|\mathcal{G}_N^{(r)}|}\sum_{G\in\mathcal{G}_N^{(r)}}\mathcal{N}_G(\rho),
\end{equation}
we can have $\Sigma(\rho)<0$. Here, $|\mathcal{G}_N^{(r)}|$ is the total number of graphs in $\mathcal{G}_N^{(r)}$. This happens for instance if $\mathcal{N}_G(\rho)=0$ for almost all $G\in\mathcal{G}_N^{(r)}$ and $\mathcal{N}_G(\rho)>0$ only for some atypical graphs. From this viewpoint, negative complexities are attributed to atypical graphs, and, more generally, to atypical samples.

Of particular interest is the point of the curve with extremal density, $\rho=\rho_\infty$, (see Fig.~\ref{fig:complexity}), which corresponds to totally {\it unfrustrated} graphs, defined as the graphs able to support a crystalline structure yielding $\rho_\infty=\rho_{\rm cryst}$; in the example of LG($k,\ell=0$) models, they correspond to graphs with no odd cycles. For these unfrustrated graphs, noted $\mathcal{G'}_N^{(r)}$, no replica symmetry breaking is expected i.e., $\mathcal{N}_G(\rho_\infty)=2=o(N)$ for $G\in\mathcal{G'}_N^{(r)}$. Since $|\mathcal{G'}_N^{(r)}|/|\mathcal{G}_N^{(r)}|$ gives the probability $P_N$ that a random graph in $\mathcal{G}_N^{(r)}$ is non frustrated, we obtain a relation between the probability $P_N$ for a sample to be non frustrated and the extremal value of the complexity $\Sigma_\infty\equiv\lim_{y\to\infty}\Sigma(y)$, 
\begin{equation}\label{eq:pN}
\lim_{N\to\infty}\frac{1}{N}\ln P_N = \Sigma_\infty.
\end{equation}
This relation constitutes the main conjecture of that paper. It is valid when $\Sigma_\infty<0$, which means that the system is typically frustrated (i.e., non satisfiable).

As shown in Fig.~\ref{fig:complexity}, the extremal density $\rho=\rho_\infty$ is reached at the terminal point $y\to\infty$ of the complexity curve; we generically expect
\begin{equation}\label{eq:phiyinfty}
\phi(y)\simeq-\rho_\infty-\frac{\Sigma_\infty}{y}\quad\quad(y\to\infty)
\end{equation}
where $\rho_\infty=\rho_{\rm rs}$ is given by a replica symmetric approach. This point basically stems from the fact that replica symmetry ignores frustration, and therefore yields the close-packing density realized on unfrustrated graphs, $\rho_\infty=\rho_{\rm cryst}$; this will be explicitly verified on the models studied below. Note finally that if unfrustrated graphs can be viewed as ideal realizations of the interior of Cayley trees, they are obviously not suitable as Bethe lattices when frustration is intended to be taken into account.

The cavity method and the interpretation of negative complexities, presented here on the example of lattice glass models, equally apply to other models where the quenched disorder may include more than just the graph structure, as for spin glass models, or may even not contain any spatial disorder at all, as for the REM. All these statements are confirmed below, where we compare for these two models, and for lattice glass models, the predictions of the 1-{\sc rsb} cavity method with some direct evaluations of large deviation properties.

\section{Negative complexity of the REM}\label{sec:rem}

The random energy model (REM) \cite{Derrida81} is renowned for being the simplest model to exhibit a 1-{\sc rsb} structure \cite{GrossMezard84} and therefore constitutes a natural testing ground. In this model, the quenched disorder is directly encoded in the $2^N$ energy levels $\{E_1,\dots,E_{2^N}\}$ which are independently taken from a Gaussian distribution
\begin{equation}
p_N(E)=\frac{1}{\sqrt{N\pi}}e^{-E^2/N}.
\end{equation}
The $N$-dependence is chosen so that the ground state is extensive, i.e., with a density
\begin{equation}
e_0=\frac{1}{N}\min\{E_1,\dots,E_{2^N}\}
\end{equation}
having a finite limit when $N\to\infty$. The complexity can directly be defined as the microcanonical entropy, through the relation
\begin{equation}\label{eq:complexityrem}
2^Np_N(Ne)\equiv e^{N\Sigma(e)},
\end{equation}
leading to $\Sigma(e)=\ln 2-e^2$. In the context of the cavity formalism, it is associated with the potential $\phi(y)=-(\ln2)/y-y/4$. The typical ground state $\overline{e_0}=-\sqrt{\ln 2}$ is obtained for $\Sigma(\overline{e_0})=0$. Note that Eq.~(\ref{eq:complexityrem}) is usually justified only when $e>e_0$ \cite{Derrida81}, but we consider here this relation for arbitrary $e$ and we will be particularly interested in the interpretation of the range where $e<e_0$.

The REM is one of the few spin glass models for which large deviations can be exactly evaluated \cite{AndreanovBarbieri03}, thanks again to the independence of the energy levels. Thus the probability $P(e)$ for $e_0$ to be $e$ is
\begin{equation}
P(e)\sim 2^N p_N(Ne)\left(\int_{Ne}^\infty dE' p_N(E')\right)^{2^N-1}.
\end{equation}
Using a saddle point approximation, we rewrite it in term of the complexity $\Sigma(e)$ as
\begin{equation}
P(e)\sim e^{N\Sigma(e)}\left(1-\frac{e^{N\Sigma(e)}}{2^N}\right)^{2^N-1}\sim e^{N\Sigma(e)-\exp (N\Sigma(e))}.
\end{equation}
Hence, we are led to distinguish the case where $\Sigma(e_0)>0$, for which we have
\begin{equation}
P(e_0>\overline{e_0})\sim e^{-\exp(N\Sigma(e_0))},
\end{equation}
from the case where $\Sigma(e_0)<0$, for which we have instead
\begin{equation}
P(e_0<\overline{e_0})\sim e^{N\Sigma(e_0)}.
\end{equation}
This last relation indicates that, in the solvable case of the REM, the whole negative branch of the complexity curve, and not only the extremal point $\Sigma_\infty$ (which is actually not defined here), can exactly be identified with the rate function for the large deviations of the ground state energy.

\section{Extreme samples and the $y\to\infty$ limit}

The next examples we consider are based on two archetypical models, the $\pm J$ spin glass and the hard-core model. In contrast with the REM, they are defined on non trivial structures, taken here as random regular graphs. For these systems, no rigorous nor conjectured exact expression is known for the ground state energy; furthermore, we are able to explicitly demonstrate a correspondence between negative complexity and large deviations only in the particular limit $y\to\infty$ where analytical calculations are amenable. For both models, we proceed in two steps, first presenting the prediction  from the 1-{\sc rsb} cavity method for the extremal complexity $\Sigma_\infty\equiv\lim_{y\to\infty}\Sigma(y)$ and, second, comparing it with the independent computation of the probability of a corresponding rare event. We will find that the conjectured relation, Eq.~(\ref{eq:pN}), is indeed exactly verified in each case.

\subsection{Spin glass model}\label{sec:spin-glass}

The Bethe spin glass model \cite{MezardParisi01} is composed of spins $s_i=\pm 1$ located on the vertices of a $(k+1)$-regular graph and interacting through the Hamiltonian
\begin{equation}
H=\sum_{\langle i,j\rangle}J_{ij}s_is_j
\end{equation}
where the couplings $J_{ij}$ are quenched variables independently chosen on each link $\langle i,j \rangle$ of the graph from the binomial distribution $\rho(J)=[\delta(J-1)+\delta(J+1)]/2$. The model has thus two sources of quenched disorder: the random regular graph spatial structure, and the random interaction couplings.

\subsubsection{The cavity approach}

The cavity method at zero temperature in the context of the spin glass model has been presented in detail by M\'ezard and Parisi \cite{MezardParisi02}, and the $y\to\infty$ limit of the potential $\phi(y)$ is found by following the principle of their calculations. On $(k+1)$-regular random graphs, we obtain
\begin{equation}
\phi(y)\simeq-\frac{k+1}{2}+\frac{k-1}{2y}\ln 2\quad\quad (y\to\infty).
\end{equation}
From the analog of Eq.~(\ref{eq:phiyinfty}), $\phi(y)\simeq \epsilon_\infty-\Sigma_\infty/y$, we extract $\Sigma_\infty=-[(k-1)/2]\ln 2$. The value $\epsilon_\infty=-(k+1)/2$ is the {\sc rs} factorized value as expected, corresponding to a system with no frustration (all links contributing to -1 in the Hamiltonian). Note that in spite of the random distribution of the couplings, the system has no local disorder, and hence can be treated in a factorized framework where all sites of the lattice are assumed to be equivalent; this is because on a tree, the typical local structure of random graphs, one can always eliminate the disorder via a gauge transformation.

The factorized 1-{\sc rsb} Ansatz is nonetheless known to be incorrect \cite{MezardParisi02}, but we shall see that it provides exact information on rare atypical unfrustrated samples. The intuitive reason is that such samples, having no frustration, do not require additional replica symmetry breaking. This will be confirmed by the direct calculation to follow, and justified in more detail through the study of the stability with respect to further {\sc rsb}, in Sec.~\ref{sec:stability}.

\subsubsection{A combinatorial approach}

In Sec.~\ref{sec:interpretation}, we claimed that $\Sigma_\infty$ is related to the probability $P_N$ that a given sample is non frustrated. We verify here that Eq.~(\ref{eq:pN}) indeed holds for the spin glass model by giving a direct evaluation of the probability $P_N$.

By definition, non frustrated samples are such that there exists a configuration $\{s^{(0)}_i\}$ of the spins with $-1=J_{ij}s^{(0)}_is^{(0)}_j$ for all $i,j$ neighbors; such samples with $J_{ij}=-s^{(0)}_is^{(0)}_j$, are called {\it Mattis models}. The total number of Mattis systems that can be defined on a graph of size $N$ is $2^N$, the number of spin configurations. Moreover, on a given $(k+1)$-regular graph of size $N$, the total number of possible realizations for the $\{J_{ij}\}$ is $2^{L_N}$ where $L_N$ is the number of links, $L_N=(k+1)N/2$. Therefore the probability to obtain a non-frustrated Mattis system from a random choice of the couplings is
\begin{equation}
P_N=2^N/2^{\frac{k+1}{2}N}=e^{-N\frac{k-1}{2}\ln 2}.
\end{equation}
Comparing with the expression for the extreme complexity, $\Sigma_\infty=-[(k-1)/2]\ln 2$, we verify that $P_N=e^{N\Sigma_\infty}$, in agreement with Eq.~(\ref{eq:pN}). The result for $P_N$ actually holds for any given $(k+1)$-regular graph, and therefore does not contain any information on atypical graphs.

\subsection{Hard-core model}\label{sec:hardcore}

The hard-core model is another simple model where the interpretation of $\Sigma_\infty$ can be checked by purely combinatorial arguments. In addition, $\Sigma_\infty$ contains in this case information on a atypical subset of the ensemble of random graphs on which the model is studied. Similar conclusions hold for more general lattice glass models and the generalization is presented in the appendix.

\subsubsection{The cavity approach}

The {\it hard-core model} is the simplest lattice glass model LG($k,\ell=0$); it has been considered previously on Erd\H{o}s-R\'enyi graphs in Ref.~\cite{WeigtHartmann01} as a model equivalent to the optimization problem called {\it minimum vertex-covering} (see Ref.~\cite{WeigtHartmann03b} for a review) and is also referred as the {\it independent set} problem in the mathematical literature, it is. In spite on the simplicity of its definition (maximally pack particles subject to the constraint that no two particles can be neighbors), no exact information is known on the phase-space structure on typical random regular or Erd\H{o}s-R\'enyi graphs; in fact, as for the spin glass model, this structure is conjectured to be of full-{\sc rsb} type.

The principle of cavity calculations in the context of lattice glasses are presented in detail in Ref.~\cite{RivoireBiroli04} for $\ell\geq 1$, and we only outline here the calculations for the $\ell=0$ case (see also Ref.~\cite{Zhou03}). Thus, following Ref.~\cite{RivoireBiroli04}, the partition function $Z_i$ for a {\it rooted-tree} with root $i$ is decomposed into a sum over two sets of configurations, $Z_i=Z^{(0)}_i+Z^{(1)}_i$, where configurations with an empty root determine $Z^{(0)}_i$, and those with an occupied root determine $Z^{(1)}_i$. A recursion in terms of the conditional partition functions of the neighbors $j=1,\dots,k$ is easily obtained,
\begin{equation}
\begin{split}
Z^{(0)}_i&=\prod_{j=1}^k\left(Z^{(0)}_j+Z^{(1)}_j\right)\\Z^{(1)}_i&=e^\mu\prod_{j=1}^kZ^{(0)}_j
\end{split}
\end{equation}
where $\mu$ is a {\it chemical potential} ($z=e^\mu$ is sometimes called the {\it activity}) weighting vertices carrying a particle. The greater $\mu>0$ is, the more occupied sites are favored; the close-packing limit corresponds to $\mu\to\infty$. A {\it cavity field} $h_i$ on site $i$ can be defined as
\begin{equation}
h_i\equiv-\frac{1}{\mu}\ln\left(\frac{Z^{(0)}_i}{Z^{(0)}_i+Z^{(1)}_i}\right).
\end{equation}
It satisfies the recursion relation
\begin{equation}
\label{eq:recursion}h_i=\hat h\left(\{h_j\}_j\right)\equiv\frac{1}{\mu}\ln\left[1+e^{\mu\left(1-\sum_j h_j\right)}\right].
\end{equation}
Because of the absence of quenched disorder and of the local homogeneity of random regular graphs, the {\sc rs} solution corresponds to a {\it liquid} phase given by the fixed point of Eq.~(\ref{eq:recursion}) with $h_i=h_{\rm liq}$ for all vertices $i$. In particular, for $\mu\to\infty$, one obtains $h_{\rm liq}\to 1/(k+1)$. This can be used to calculate densities \cite{RivoireBiroli04}, yielding $\rho_{\rm rs}(\mu)\to 1/2$ as $\mu\to\infty$. This value corresponds to a crystalline close-packing $\rho_{\rm rs}=\rho_{\rm cryst}$ and is certainly an overestimation of the random close-packing value $\rho_{\rm rcp}$, as already discussed in Sec.~\ref{sec:complexity}. In fact, the liquid solution can be shown to become unstable beyond a certain chemical potential $\mu_c$.

The 1-{\sc rsb} Ansatz provides a better approximation that takes into account some of the frustration effects by allowing the phase space to break into states i.e., clusters of configurations. In the $\mu\to\infty$ limit, Eq.~(\ref{eq:recursion}) becomes
\begin{equation}
h_i=\max\left(0,1-\sum_{j=1}^k h_j\right).
\end{equation}
The 1-{\sc rsb} order parameter is then a distribution $P(h)$ over cavity fields associated with different states that can be written as $P(h)=p\delta(h)+(1-p)\delta(h-1)$. It is described by a single real $p\in [0,1]$ which satisfies the self-consistent equation
\begin{equation}
\label{eq:1rsb}p=\frac{1-p^k}{1+p^k(e^y-1)}.
\end{equation}
To obtain $\rho_{\rm rcp}$, the parameter $y$ must be chosen to maximize $\phi(y)$, here given by
\begin{equation}
\phi(y)=-\frac{1}{y}\left[\ln\left(1+p^{k+1}(e^y-1)\right)-\frac{k+1}{2}\ln\left(1+(1-p)^2(e^{-y}-1)\right)\right].
\end{equation}
On the other hand, if one focuses on the $y\to\infty$ limit, one finds that $p\sim e^{-y/(k+1)}$ and
\begin{equation}
\phi(y)\simeq-\frac{1}{2}+\frac{k-1}{2y}\ln 2\quad (y\to\infty).
\end{equation}
Referring to Eq.~(\ref{eq:phiyinfty}), we read off $\rho_\infty=1/2$ and $\Sigma_\infty=-\frac{k-1}{2}\ln 2$. As anticipated, we obtain $\rho_\infty=\rho_{\rm rs}=\rho_{\rm cryst}=1/2$, the maximum density on a graph with no frustrating loops. (This {\sc rs} value is different from the non-factorized one that can be calculated at $y=0$.) For $\ell=0$, unfrustrated graphs are graphs with no odd cycles and are called {\it bipartite} or {\it bicolorable} graphs.

\subsubsection{A combinatorial approach}\label{sec:combinatorics}

According to Eq.~(\ref{eq:pN}), the complexity $\Sigma_\infty<0$ is related to the probability for a random $r$-regular graph ($r=k+1$) to be bicolorable. In order to check this probabilistic interpretation of $\Sigma_\infty$, we present an independent derivation of the probability of non frustrated samples based on the asymptotic estimation of the probability $P_N(r)=|\mathcal{G'}_N^{(r)}|/|\mathcal{G}_N^{(r)}|$ that a random $r$-regular admits $\rho_\infty=1/2$ as maximum density i.e., is bipartite. As in the cavity method, the idea is to study the system under addition of one vertex (in fact adding {\it two} vertices at a time turns out to be more appropriate). The idea is to compute $\lambda_2(r)$ such that $P_{N+2}(r)\sim\lambda_2(r) P_N(r)$ (asymptotically in $N$), in order to write $P_N(r)\sim\lambda_2(r)^{N/2}$ and verify that we indeed have $\Sigma_\infty=\frac{1}{2}\ln \lambda_2(r)$ i.e., $e^{N\Sigma_\infty}\sim P_N(r)$ as predicted in Eq.~(\ref{eq:pN}).
\begin{figure}
%\centering\resizebox{0.5\textwidth}{!}{\epsfig{file=ntonp2.eps}}
\centering\epsfxsize=8cm \epsfbox{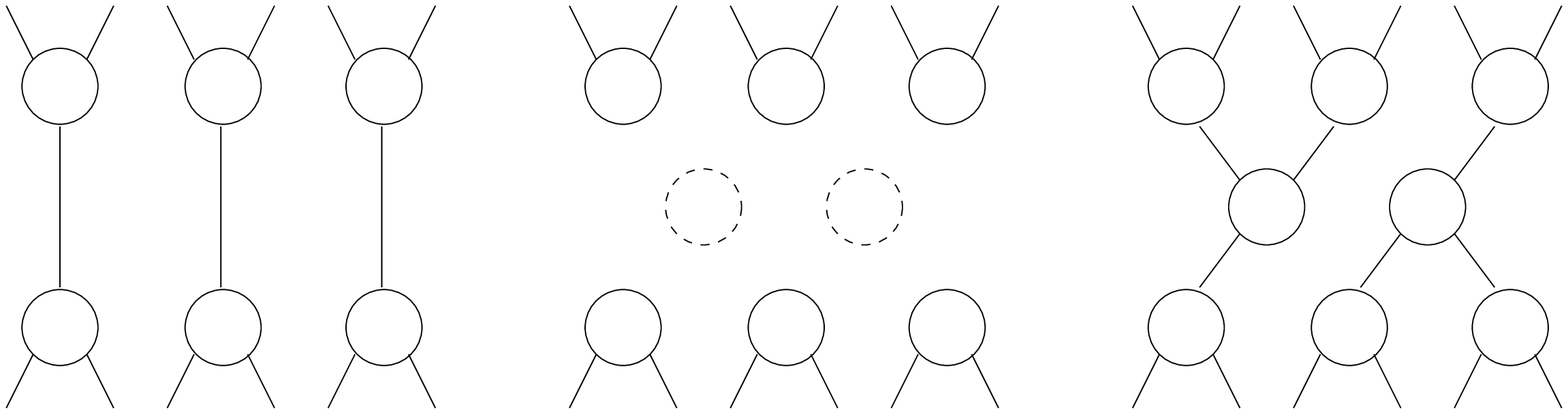}
\caption{\small Addition of two vertices to build a $r$-regular graph of size $N+2$ from one of size $N$, illustrated here with $r=k+1=3$. First, $r=3$ edges are chosen at random and cut. Second, two new vertices are connected to the $2r=6$ vertices left with one missing edge.}\label{fig:total}
\end{figure}

To do so, we successively estimate the total number $|\mathcal{G}_N^{(r)}|$ of random regular graph with size $N$ and degree $k+1=r$, and the number $|\mathcal{G'}_N^{(r)}|$ of unfrustrated (bipartite) random regular graphs of size $N$. For convenience, the graphs we count here are {\it unlabelled} graphs i.e., defined up to an isomorphism of the vertices, but, as discussed below, the distinction with labelled graphs, is actually irrelevant in the large $N$ asymptotics. Firstly, to relate $|\mathcal{G}_N^{(r)}|$ and $|\mathcal{G}_{N+2}^{(r)}|$, we construct a graph with $N+2$ vertices from one with $N$ vertices. One way to add two vertices while preserving a constant degree $r$ is to remove $r$ edges and connect two new vertices to the $2r$ vertices having one edge removed, as shown in Fig.~\ref{fig:total}. The total number of edges is $rN/2$, so we have $C(rN/2,r)$ choices when deciding the edges to cut, where $C(n,p)\equiv n!/[p!(n-p)!]$ denotes the binomial coefficient giving the number of ways to partition $n$ elements into groups of $p$. For the recombination, we have $R(r,2)$ choices, where $R(n,p)\equiv(np)!/[p!(n!)^p]$ corresponds to the number of ways to gather $np$ objects into $p$ indiscernible packs of $n$ each. This way, all random regular graphs with $N+2$ vertices are constructed from those with $N$. However, we did some overcounting: each graph with $N+2$ vertices can be obtained from a certain number of different graphs of size $N$. To evaluate that number, we consider a graph with $N+2$ vertices: any pair of vertices can correspond to the two newly added vertices, and there are $C(N+2,2)$ such pairs. Given a pair of vertices, the graph can result from $R(2,r)$ different graphs of size $N$: they are found by deleting the edges of the pair of vertices and recombining in all the possible ways the vertices left with one missing edge. Finally, we get
\begin{equation}\label{eq:N-N+2}
|\mathcal{G}_{N+2}^{(r)}|\sim\frac{C(rN/2,r)R(r,2)}{C(N+2,2)R(2,r)}|\mathcal{G}_N^{(r)}|.
\end{equation}
This is only asymptotically correct, since we tacitly assumed for instance that two edges picked at random do not share a common vertex.
\begin{figure}
%\centering\resizebox{0.5\textwidth}{!}{\epsfig{file=ntonp2bip.eps}}
\centering\epsfxsize=8cm \epsfbox{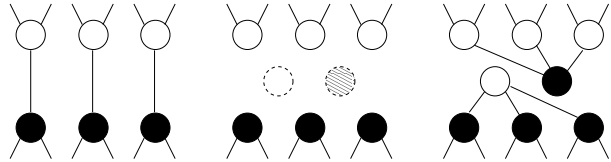}
\caption{\small Addition of two vertices to build a $r$-regular {\it bipartite} graph of size $N+2$ from one of size $N$. For a bipartite graph, each vertex can be colored either in white or in black, such that no identically colored vertices are adjacent. The $r=3$ edges chosen at random to be cut thus necessarily connect two vertices with different colors. To obtain again a bipartite graph when adding two vertices, the two new vertices must be of different colors, and no choice is left to connect them to the old ones.} \label{fig:bipartite}
\end{figure}
Next, we relate $|\mathcal{G'}_N^{(r)}|$ and $|\mathcal{G'}_{N+2}^{(r)}|$ by starting from a bipartite graph of size $N$ and by constructing one of size $N+2$. By definition, it is possible to color each vertex in white (empty) or black (occupied by a particle), with no edge between two vertices sharing the same color. If we add two vertices, there must necessarily be one white and one black. Again, deleting $r$ edges can be done in $C(rN/2,r)$ ways, but this time there is only one way to connect the two new vertices with the old ones (see Fig.~\ref{fig:bipartite}). To evaluate the degeneracy, we have to consider one of the $[(N+2)/2]^2$ pairs of white and black vertices and count the number of ways to reobtain a bipartite graph of size $N$ if they are removed: since we are left with $r$ vertices of each color, this number is $r!$, the number of mappings from the white to the black vertices. Thus, we get
\begin{equation}\label{eq:N-N+2prime}
|\mathcal{G'}_{N+2}^{(r)}|\sim\frac{C(rN/2,r)}{[(N+2)/2]^2r!}|\mathcal{G'}_N^{(r)}|.\end{equation}
We are then in position to calculate
\begin{equation}
\lambda_2(r)\equiv\frac{P_{N+2}(r)}{P_N(r)}=\frac{|\mathcal{G}_N^{(r)}|}{|\mathcal{G}_{N+2}^{(r)}|}\frac{|\mathcal{G'}_{N+2}^{(r)}|}{|\mathcal{G'}_N^{(r)}|}\sim 2^{-(r-2)}=2^{-(k-1)},
\end{equation}
in agreement with our expectations since it yields $\Sigma_\infty=\frac{1}{2}\ln\lambda_2$. The complexity $\Sigma_\infty$ thus indeed gives the probability $P_N\sim2^{-N(r/2-1)}$ for a random $r$-regular graph to be bipartite.

\subsubsection{Connections with graph theory}

Interestingly, Eq.~(\ref{eq:N-N+2}) also gives an estimation of the total number of unlabelled random regular graphs, a quantity which, in contrast to the case of the Erd\H{o}s-R\'enyi class, is not straightforwardly obtained \cite{Bollobas01}. We calculated indeed $|\mathcal{G}_{N+2}^{(r)}|/|\mathcal{G}_N^{(r)}|\sim r^rN^{r-2}/(r!)^2\equiv\kappa_r(N)$, therefore
\begin{equation}
\label{eq:moi}|\mathcal{G}_N^{(r)}|\sim\prod_{n=1}^N\kappa_r(n)^{1/2}\sim\frac{r^{Nr/2}(N!)^{r/2-1}}{(r!)^N}\equiv V_r(N).
\end{equation}
This can be compared to known mathematical results. The mathematical literature is mostly concerned with {\it labelled} random graphs, having distinguishable vertices. However, since the automorphism group of a random graph is in fact typically trivial \cite{Bollobas01}, the number of unlabelled graphs can be obtained from the number of labelled graphs merely by dividing with $N!$. Thus, the number $U_r(N)$ of unlabelled regular graphs is known to satisfy \cite{Bollobas01, Wormald99},
\begin{equation}\label{eq:math}
|\mathcal{G}_N^{(r)}|\sim\frac{(rN)!e^{-(r^2-1)/4}}{(rN/2)!2^{rN/2}(r!)^NN!}\equiv U_r(N).
\end{equation}
Formulae (\ref{eq:moi}) and (\ref{eq:math}) are asymptotically equivalent, in the sense that $\ln U_r(N)=\ln V_r(N)+o(N)$. In Eq.~(\ref{eq:math}), the precise factor $e^{-(r^2-1)/4}$ reflects the fact that $U_r(N)$ includes only {\it simple} graphs i.e., graphs with no loop of size one or multiple edges \cite{Wormald99}, a detail inaccessible to the too crude asymptotic estimations of the cavity method. Similarly, the number of bipartite graphs as estimated from Eq.~(\ref{eq:N-N+2prime}) is consistent with an asymptotic estimation, due to O'Neil \cite{ONeil69}, of the number of bicolorable {\it labelled} $r$-regular on $N$ vertices \cite{Wormald99},
\begin{equation}\label{eq:ONeil}
B_r(N)=\frac{(rN/2)!e^{-(r-1)^2/2}}{(r!)^N}.
\end{equation}
Here again, the number of {\it unlabelled} bipartite graphs can be obtained by dividing Eq.~(\ref{eq:ONeil}) with the expected number $A_r(N)$ of automorphisms of labelled bipartite $r$-regular graphs of size $N$ \cite{Wormald99}. Since the labelling of a bipartite graph is $1,2,\dots,N/2$ for the vertices of each color, one has to use $A_r(N)=[(N/2)!]^2$ to verify that Eqs.~(\ref{eq:N-N+2prime}) and (\ref{eq:ONeil}) are indeed consistent. If the same labelling is kept for bipartite and non-bipartite graphs, the triviality of the automorphism groups indicates that the asymptotic probability $P_N(r)$ can be indifferently obtained as a ratio of labelled or unlabelled graphs. Therefore, the distinction between distinguishable and indistinguishable vertices, while crucial when counting small graphs, is irrelevant in the asymptotic regime where the cavity method operates. Coming back to atypical graphs, we already noted that the cavity method on hard-core models yields the probability for a random $r$-regular graph ($r=k+1$) to be bicolorable,
\begin{equation}P_N(r)\sim 2^{-\left(\frac{r}{2}-1\right)N}.
\end{equation}% note the r=2 case
Similar conclusions should apply to other random graph ensembles; thus for Erd\H{o}s-R\'enyi graphs with mean-connectivity $\gamma$, the result of Zhou \cite{Zhou03}, $\Sigma_\infty=-\pi^2(\ln\gamma-1)^2/16\gamma$, can be used to infer
\begin{equation}
P_N(\gamma)\sim e^{-\frac{\pi^2(\ln\gamma-1)^2}{16\gamma}N}.
\end{equation}
Note that this relation holds only for $\gamma>e$, since low connectivities $\gamma<e$ correspond here to a {\sc rs} regime where graphs are typically i.e., {\it asymptotically almost surely}, bicolorable, that is $P_N(\gamma\leq e)=1$. Other ensembles on which the hard-core model has been studied include random graphs with arbitrary degree distribution \cite{VazquezWeigt03} and random graphs with short local loops \cite{WeigtHartmann03}.

\subsection{Validity of the {\sc 1-rsb} approach}\label{sec:stability}

The previous discussion resorted to a 1-{\sc rsb} Ansatz which is known to be wrong for the spin glass model \cite{MezardParisi01}, and to correctly describe LG($k,\ell$) models only for a restricted set of the parameters $\ell$ and $k$, as discussed in Ref.~\cite{RivoireBiroli04} for $\ell\geq 1$. In general, a 1-{\sc rsb} Ansatz can undergo two kinds of instabilities \cite{MontanariRicci03}: one of a first kind for $y>y^{(1)}_c$ and of a second kind for $y<y^{(2)}_c$. For the LG($k,\ell=0$) models, one finds $y^*<y^{(2)}_c<y^{(1)}_c$ if $k=2$, in which case only a restricted negative part of the complexity curve is stable, while for all other values of $k\geq 3$ studied, $y^{(1)}_c<y^{(2)}_c$ and no part of the complexity curve is stable to further breaking of the replica symmetry. Therefore, in the case of the hard-core model too, our conclusions based on 1-{\sc rsb} calculations are questionable. In this respect, the observed agreement, for {\it all} models studied, between the 1-{\sc rsb} complexity at infinite $y$ and the independent counting arguments looks quite puzzling.

This suggests that the 1-{\sc rsb} approximation in fact provides exact predictions in the $y\to\infty$ limit, thus generalizing the observation that the {\sc rs} approximation also carries exact information on this limit, in the sense that $\phi(y=\infty)=-\rho_\infty=-\rho_{\rm rs}$. We therefore propose that the 1-{\sc rsb} complexity $\Sigma_\infty$ gives the {\it exact} first correction to an expansion of a {\it full-{\sc rsb}} potential $\phi^{(\infty)}$ around its minimum i.e., $\phi^{(\infty
)}(y)=-\rho_\infty-\Sigma_\infty/y+O(1/y)$ for $y\to\infty$ with $\Sigma_\infty$ obtained from the 1-{\sc rsb} approximation.

To check this point explicitly, we consider the stability of the 1-{\sc rsb} Ansatz in the framework of 2-{\sc rsb} cavity equations. For sake of concreteness, we restrict ourselves to LG$(k,\ell=0)$ models, but the conclusions should hold for more general models. In this context, the order parameter is a distribution $\mathcal{Q}[P]$ over cavity field distributions $P_a(h)$ which are in fact described, as noted in Sec.~\ref{sec:hardcore}, with a single real $p_a$ such that $P_a(h)=p_a\delta(h)+\left(1-p_a\right)\delta(h-1)$. The cavity equation at the 2-{\sc rsb} level reads
\begin{equation}\label{eq:2rsb}
\mathcal{Q}(p_0)=\frac{1}{Z}\int\prod_{j=1}^kdp_j \mathcal{Q}(p_j)\delta\left(p_0-\frac{1-\prod_jp_j}{1+(e^{y_2}-1)\prod_j p_j}\right)\left(1+(e^{y_2}-1)\prod_{j=1}^k p_j\right)^{y_1/y_2},
\end{equation}
with $y_1\leq y_2$, where $y_2$ is associated with the smaller clusters, the states, and $y_1$ with the larger clusters, the families of states. Equation (\ref{eq:2rsb}) admits as an exact solution the peaked distribution $\mathcal{Q}(p)=q\delta(p-1)+(1-q)\delta(p)$. It corresponds in fact to the embedding of the 1-{\sc rsb} Ansatz into a 2-{\sc rsb} framework, which is always possible. More precisely, this solution satisfies the analog of Eq.~(\ref{eq:1rsb}) with $q$ and $y_1$ respectively playing the role of the 1-{\sc rsb} parameters $p$ and $y$ (and $y_2$ playing no role).

Let us now consider the expansion of the 2-{\sc rsb} potential $\phi^{(2)}(y_1,y_2)$ around its minimum $-\rho_\infty$, reached for $y_1=y_2=\infty$. The first correcting term is expected to come from $y_1$, since $y_1\leq y_2$. As $y_2$ is increased, one can see from Eq.~(\ref{eq:2rsb}), that $\mathcal{Q}$ get peaked on the integers 0 and 1 and for $y_2\to\infty$ is compelled to reproduce a 1-{\sc rsb} solution (that can be checked numerically by population dynamics). This means that $\phi^{(2)}(y_1,y_2=\infty)=\phi(y_1)$ is given by the 1-{\sc rsb} Ansatz. We limited ourselves here to 2-{\sc rsb}, but by virtue of the recursiveness of the {\sc rsb} scheme, the same feature repeats itself at each successive level of the hierarchy, yielding $\phi^{(m)}(y_1,y_2=\infty,\dots,y_m=\infty)=\phi^{(1)}(y_1)\equiv\phi(y_1)$ for arbitrary $m\geq 1$. This entails that $\Sigma_\infty$ as computed within a 1-{\sc rsb} framework indeed corresponds to the first correction in $1/y_1$ of the $m$-{\sc rsb} potential $\phi^{(m)}$ in the vicinity of its minimum $-\rho_\infty$, and, by extension to $m\to\infty$, of the full-{\sc rsb} potential $\phi^{(\infty)}$.

In the case of the spin glass or the LG($k,\ell=0$) hard-core models, deviations from 1-{\sc rsb} predictions are expected as soon as $y$ is finite, since the 1-{\sc rsb} Ansatz is then unstable while, in more favorable cases, such as for instance $\ell=1$ and $k=2, 3$, the whole negative part of the complexity curve is correctly described by an 1-{\sc rsb} Ansatz \cite{RivoireBiroli04}.

\section{Conclusion}\label{sec:conclusion}

The negative branch of the complexity curve as obtained from the cavity method is usually disregarded as an irrelevant, unphysical feature. We argued here that its extremal value, $\Sigma_\infty$, is related to large deviation properties of the disorder realization, i.e., to the exponentially small (with respect to the size of the system) probabilities of observing atypical samples. We explicitly checked this relation in several particular cases where exact computations could be done.

As an application, we showed that when the quenched disorder lies in random graph spatial structure, 1-{\sc rsb} calculations provide an efficient tool to evaluate asymptotic probabilities of atypical graphs. The simplicity of the method can be appreciated when comparing it to equivalent combinatorial computations. These combinatorial arguments are however interesting not only because they validate the proposed interpretation, but also because they constitute a new situation where 1-{\sc rsb} predictions are confronted with alternative derivations.

The focus was mainly put on models defined on random regular graphs, but the same principles are expected to apply without additional difficulties to any other model defined on arbitrary ensembles of random dilute graphs for which $\Sigma_\infty<0$. This includes notably various optimization problems on Erd\H{o}s-R\'enyi (hyper)graphs where cavity calculations are already available \cite{MezardZecchina02, MuletPagnani02, Zhou03}. In this context, recognizing that large deviation properties can be obtained from well known statistical physics techniques could be a first step toward a reconciliation between the {\it worst-case} (atypical) viewpoint of computer scientists and the {\it typical-case} viewpoint on which most of ``equilibrium''  physics have focused up to now.

Finally, besides these potential applications to random graphs or optimization problems, we note that large deviations are already known to be deeply connected to statistical mechanics, and, more specifically, to the concept of entropy \cite{Ellis85}. This work suggests that replica theory is also intimately related to the large deviations of the free energy with respect to a quenched disorder. The open problem of extending the present approach to the computation of the whole free energy probability distribution function beyond the extremal case of totally non frustrated samples may unravel additional connections with replica theory.

\subsection*{Acknowledgements}

I thank Olivier Martin and Marc M\'ezard for discussions and a critical reading of the manuscript. This work was supported in part by the European Community's Human Potential Programme under contracts HPRN-2002-00307 (DYGLAGEMEM) and HPRN-CT-2002-00319 (STIPCO).

\appendix

\section{Appendix: Lattice glass models}

This appendix is devoted to the extension of the methods and results of Sec.~\ref{sec:hardcore} on the hard-core model to lattice glass models with $\ell>0$. It leads to further support in favor of the interpretations of $\rho_\infty$ and $\Sigma_\infty$, and to some additional results on atypical graphs.

\subsection{The cavity approach}\label{sec:latticeglass}

The $y\to\infty$ limit for the 1-{\sc rsb} solution of LG($k,\ell$) models in the close-packing limit ($\mu\to\infty$) yields

\begin{equation}
\begin{split}\label{eq:rhoinftykl}
\rho_\infty=&\frac{k+1}{2k-\ell+2}\\
\Sigma_\infty=&-\frac{k-1}{2}\ln(2k-\ell+2)+\frac{k^2-k\ell-1}{2k-\ell+2}\ln(k-\ell+1)\\
&+\frac{k+1}{2}\frac{\ell}{2k-\ell+2}\ln\ell+\frac{k+1}{2k-\ell+2}\ln\left(\frac{k!}{\ell!(k-\ell)!}\right).
\end{split}
\end{equation}
\begin{figure}
%\centering\resizebox{0.2\textwidth}{!}{\epsfig{file=rhoinfty.eps}}
\centering\epsfxsize=4cm \epsfbox{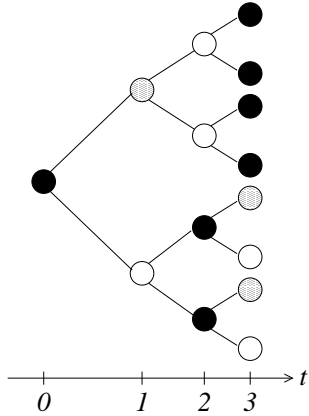}
\caption{\small Pictorial representation of the rules (\ref{eq:rule}) used to determine $\rho_\infty$ in LG($k,\ell$) models, here for $k=2$ and $\ell=1$. The state of a vertex at generation $t+1$ is determined by the state of its parent at generation $t$. If the parent is empty (noted '0' in the text and colored in white in the figure), its children are filled with a particle; they define the state '1' (in black) of occupied vertices with an empty parent. Such states generate children in state '2' (in grey), which are occupied vertices with an occupied parent too. Due to the constraint $\ell=1$, they cannot have occupied children and generate therefore empty vertices (state 0). The resulting population dynamics is described by the rate equations (\ref{eq:rate}).} \label{fig:rhoinfty}
\end{figure}
As before, $\rho_\infty$ is indeed the (factorized) {\sc rs} value, $\rho_\infty=\rho_{\rm rs}$. The correspondence with the close-packing density on unfrustrated graphs can independently be demonstrated by building up a crystal in the following way: we start on a vertex and assume that no loop prevents us from choosing the best possible children at each successive generation. Noting $'0'$ when a vertex is empty,$'1'$ when it is occupied but its ancestor is empty and $'2'$ when both it and its ancestor are occupied, the rule, as illustrated in Fig.~\ref{fig:rhoinfty}, reads
\begin{equation}\label{eq:rule}
\begin{split}0&\to\underbrace{1+\dots+1}_k,\\
1&\to\underbrace{2+\dots+2}_\ell+\underbrace{0+\dots+0}_{k-\ell},\\
2&\to\underbrace{2+\dots+2}_{\ell-1}+\underbrace{0+\dots+0}_{k-\ell+1}.
\end{split}
\end{equation}
One deduces the number $N_i$ of sites of each kind $i=0,1,2$ at generation $t+1$ from that numbers at generation $t$:
\begin{equation}\label{eq:rate}
\begin{split}
N_0(t+1)&=(k-\ell)N_1(t)+(k-\ell+1)N_2(t),\\
N_1(t+1)&=kN_0(t),\\N_2(t+1)&=\ell N_1(t)+(\ell-1)N_2(t).
\end{split}
\end{equation}
Fixed points of the form $x_1=N_1/N_0$ and $x_2=N_2/N_0$ are found to satisfy the equation $x_1^3+(k-\ell)x_1^2+(\ell-1)x_1-k=0$ whose only positive root is $x_1=1$, giving $x_2=\ell/(k-\ell+1)$ and
\begin{equation}
\rho=\frac{x_1+x_2}{1+x_1+x_2}=\frac{k+1}{2k-\ell+2},
\end{equation}
which coincides as expected with the $\rho_\infty$ given by Eq.~(\ref{eq:rhoinftykl}).

\subsection{Combinatorial approach}

Here again $\Sigma_\infty<0$ can be derived by purely combinatorial means, and we present such calculations for the case $\ell=1$. In this context ($\ell=1$), the final result can be interpreted as the probability that a $(k+1)$-regular graph has a close-packing of dimers, where dimers lie on the edges and exclude the neighboring {\it and} next-neighboring edges to be occupied. More generally, $P_N=\exp[N\Sigma_\infty]$ represents the probability that a random $(k+1)$-regular graph can support the extremal density $\rho_\infty$, which means for general $\ell$ that it is possible to assign particles on some of the vertices such that each vertex is either empty with $k+1$ occupied neighbors, or occupied with exactly $\ell$ occupied neighbors.

The calculation follows the same principles as the one exposed in Sec.~\ref{sec:hardcore} for the $\ell=0$ constraint. Unfrustrated graphs here correspond to graphs where each vertex either is occupied with exactly one neighbors or is empty with $r=k+1$ occupied neighbors. To keep the maximal density $\rho_\infty=(k+1)/(2k+1)$, it is here necessary to add at each step a number of vertices which is a multiple of $2k+1$; since on the other hand this number has to be even for the construction to be doable, we add $2(2k+1)$ vertices at a time. The generalization of Eq.~(\ref{eq:N-N+2}) for the addition of $2p$ instead of only $2$ vertices (with in view $p=2k+1$) is
\begin{equation}
\frac{|\mathcal{G}_{N+2p}^{(k+1)}|}{|\mathcal{G}_N^{(k+1)}|}\sim\frac{C(\frac{k+1}{2}N,p(k+1))R(k+1,2p)}{C(N+2p,2p)R(2,p(k+1))}
\end{equation}
where, as before, the calculations are done for unlabelled graphs, and we note $C(n,p)\equiv n!/[p!(n-p)!]$, $R(n,p)\equiv(np)!/[p!(n!)^p]$. The hard part is to relate the number of unfrustrated graph $|\mathcal{G'}_{N+2(2k+1)}^{(k+1)}|$ of size$N+2(2k+1)$ to that of size $N$. Starting from an unfrustrated graph of size $N$, we can distinguish empty vertices, of density $\rho_0=k/(2k+1)$, and occupied vertices, of density $\rho_1=(k+1)/(2k+1)$. We also need to consider the two kinds of edges, those relating one empty and one occupied vertex (edges of "type 1"), whose density is $\pi_1=2k/(2k+1)$ and those relating two occupied vertices (edges of "type 2"), whose density is $\pi_2=1/(2k+1)$. Among the $2(2k+1)$ vertices added, $2(k+1)$ must be occupied and $2k$ empty for $\rho_\infty$ to be maintained. If we want to be able to connect them to preexisting vertices with one removed edges, the edges to delete must be chosen carefully: it can be seen that we have to delete exactly $2k(k+1)$ edges of type 1 and $(k+1)$ of type 2. Since there a total number of $\pi_iN(k+1)/2$ edges of each type $i=1,2$, the number $A_1(k,N)$ of admissible choices for the edge deletion is
\begin{equation}
A_1(k,N)=C\left(\pi_1N(k+1)/2,2k(k+1)\right)C\left(\pi_2N(k+1)/2,k+1\right).
\end{equation}
To connect the new vertices, we have $R(k+1,2k)$ choices for the new empty vertices, and $[2(k+1)]!R(k,2(k+1))$ choices the new occupied ones ; the total number of choices is therefore
\begin{equation}
A_2(k,N)=[2(k+1)]!R(k,2(k+1))R(k+1,2k).
\end{equation}
Now we evaluate the degeneracy of the construction, that is the number of graphs of size $N$ leading to the same graph of size $N+2(2k+1)$. As in the $\ell=0$ case, we do so by starting from an unfrustrated graph of size $N+2(2k+1)$ and counting the number of unfrustrated graphs of size $N$ obtained by deleting $2(2k+1)$ vertices. Here again, the choice of the vertices to remove is constrained: we have to delete $2(k+1)$ occupied vertices out of the $\rho_1[N+2(2k+1)]$ ones, and $2k$ empty vertices out of the $\rho_0[N+2(2k+1)]$ ones, resulting in a number of choices
\begin{equation}
A_3(k,N)=C(\rho_1[N+2(2k+1)),2(k+1))C(\rho_0[N+2(2k+1)),2k).
\end{equation}
Finally, we need the number of ways $A_4(k,N)$ to add edges to reform an unfrustrated regular graph of size $N$. The $2(k+1)$ occupied vertices left without neighbor must be recombined together, giving a factor $R(2,k+1)$ while the $2k(k+1)$ occupied ones with already one neighbor must be matched with the $2k(k+1)$ empty vertices, yielding a factor $[2k(k+1)]!$, so that
\begin{equation}
A_4(k,N)=[2k(k+1)]!R(2,k+1).
\end{equation}
This leads to the relation
\begin{equation}
\frac{|\mathcal{G'}_{N+2(2k+1)}^{(k+1)}|}{|\mathcal{G'}_N^{(k+1}|}\sim\frac{A_1(k,N)A_2(k,N)}{A_3(k,N)A_4(k,N)}
\end{equation}
from which one deduces an asymptotical estimation of
\begin{equation}
\lambda_{2(2k+1)}(N)\equiv\frac{P_{N+2(2k+1)}}{P_N}=\frac{|\mathcal{G'}_{N+2(2k+1)}|}{|\mathcal{G}_{N+2(2k+1)}^{(k+1)}|}\frac{|\mathcal{G}_N^{(k+1)}|}{|\mathcal{G'}_N^{(k+1)}|}.
\end{equation}
The result obtained is
\begin{equation}
\frac{1}{2(2k+1)}\ln \lambda_{2(2k+1)}(N)\sim\frac{k^2}{2k+1}\ln k-\frac{k-1}{2}\ln (2k+1)=\Sigma_\infty
\end{equation}
in precise agreement with the interpretation of $\Sigma_\infty$ given in the text.

\bibliographystyle{unsrt}
\bibliography{graphs,glasses,references}

\end{document}